\documentclass[10pt]{article}
\usepackage[utf8]{inputenc}
\usepackage{graphicx}
\usepackage[utf8]{inputenc}
\usepackage{amsmath}
\usepackage{amsfonts}
\usepackage{amssymb}
\usepackage{graphicx}

\usepackage{amsmath, physics, siunitx}

\usepackage{epstopdf}
\AppendGraphicsExtensions{.tif}

\usepackage{setspace}
\usepackage{authblk}
\usepackage{caption}
\usepackage{float}
\usepackage{cite}
\usepackage[left=2.0cm,right=2.0cm,top=2cm,bottom=2cm]{geometry}
\usepackage{color}
\usepackage{color,soul}
\usepackage{multicol}
\usepackage{subcaption}

\title{\textbf{Conventional Scintillation Statistics with Turbulence Impacted Coupled Dipole Oscillation}}

\author[1,*]{Shouvik Sadhukhan}

\author[2]{C. S. Narayanamurthy}

\affil[1, 2]{\small{Applied and Adaptive Optics Laboratory, Department of Physics, Indian Institute of Space Science and Technology (IIST), P.O: Valiamala, Trivandrum - 695547, State: Kerala; India}}
\affil[1]{\small{Email: shouvikphysics1996@gmail.com}}
\affil[2]{\small{Email: naamu.s@gmail.com}}
\affil[*]{\small{Corresponding Author Email: shouvikphysics1996@gmail.com}}

\begin{document}
\maketitle

\begin{abstract}
We investigate the propagation of optical fields through polymethyl methacrylate (PMMA) rods under atmospheric turbulence conditions, employing a generalized Lorentz dipole oscillator model with nonlinear restoring forces and dipole-dipole coupling. The theoretical framework incorporates second- and third-order anharmonic terms ($\beta_i|r_i|r_i$ and $\alpha_i|r_i|^2r_i$) alongside dyadic Green's function-mediated coupling between localized dipoles. Gradient forces arising from spatially non-uniform field distributions and Lorentz force perturbations are incorporated through d'Alembert's principle, revealing an effective inertia mechanism that opposes rapid field redistribution. Modal diagonalization demonstrates that synchronized dipole oscillations can compensate turbulence-induced wavefront distortions, with the perturbation force $\delta F_{\text{Pert}}(t) = F'_{\text{Inertia}} - F_{\text{Inertia}}$ governing the compensation efficacy. Experimental verification employs a pseudo-random phase plate (PRPP) generating Kolmogorov-spectrum turbulence, with 200 frames recorded across four configurations: baseline, turbulence-only, and turbulence with one or two PMMA rods. Statistical analysis quantifies scintillation index variations. Results indicate that dipole-dipole coupling energy transitions enable partial turbulence compensation when stronger suppression observed for longer propagation paths through increased synchronization.\\

\end{abstract}

\section{Introduction}
The propagation of coherent optical fields through turbulent media remains a fundamental challenge in atmospheric optics, free-space optical communication, and remote sensing applications. Atmospheric turbulence induces random fluctuations in the refractive index, leading to wavefront distortions, beam wander, and intensity scintillation that severely degrade system performance. Understanding and mitigating these effects requires a comprehensive theoretical framework that captures both the stochastic nature of turbulence and the complex light-matter interactions within propagating media. When optical beams traverse dielectric materials under turbulent conditions, the interplay between electromagnetic field dynamics and molecular dipole oscillations introduces additional degrees of freedom that can either amplify or suppress turbulence-induced degradation. The Lorentz oscillator model, which describes bound electron dynamics in response to external electromagnetic fields, provides a natural starting point for analyzing these phenomena. However, conventional linear treatments are insufficient when dealing with materials exhibiting significant nonlinear optical responses or when dipole-dipole coupling becomes non-negligible. This necessitates an extended theoretical approach incorporating anharmonic restoring forces, electromagnetic coupling between localized dipoles, and gradient-induced optical forces. \cite{1,2,3,4,5,6,7,8,9,10,11,12,13,14,15,16,17,18,19,20,21,22,23,24,25}\par

Extensive research has been conducted on the statistical properties of optical beams propagating through random media. Korotkova and colleagues have made significant contributions to understanding intensity fluctuations and scintillation behavior of electromagnetic beams in turbulent atmospheres \cite{2,3,4,5}. Their work on scintillation index calculations for stochastic electromagnetic beams \cite{5} and changes in instantaneous Stokes parameters \cite{3,4} established fundamental frameworks for characterizing beam propagation statistics. Further investigations into oceanic turbulence effects on polarization \cite{7} and the development of Stokes-Mueller correlation calculus \cite{8} have expanded the understanding of electromagnetic wave propagation in complex random media. Recent advances in absorption and scattering in natural waters \cite{9}, along with the introduction of the Poincaré sphere of electromagnetic spatial coherence \cite{10}, have provided powerful tools for analyzing polarization and coherence properties. The unified matrix representation for spin and orbital angular momentum in partially coherent beams \cite{14} and the coherence Poincaré sphere formulation \cite{15} have further enriched the mathematical description of optical field propagation through turbulent media. These statistical approaches, while comprehensive in characterizing beam propagation, generally treat the propagating medium as a passive perturbative element without explicitly accounting for the microscopic dipole dynamics and energy exchange mechanisms within the material. \cite{26,27,28,29,30,31,32,33,34,35,36,37,38,39,40,41,42,43,44,45,46,47,48,49,50}\par

The microscopic understanding of light-matter interaction through dipole dynamics has received considerable attention in nanophotonics and metamaterials research. The investigation of dipole emission near dielectric metasurfaces \cite{16} and the macroscopic quantum electrodynamics framework \cite{17} provide rigorous theoretical foundations for electromagnetic interactions in structured media. Classical electromagnetic scattering theories, including extensions of Green's function methods \cite{18,21,22}, have been instrumental in understanding field propagation in polarizable backgrounds. The discrete dipole approximation has proven particularly valuable for predicting scattering from complicated metamaterials \cite{25}, while studies on dipole polarizability of time-varying particles \cite{26} have revealed rich dynamical behavior. Fano's seminal work on normal modes of coupled oscillators \cite{27} laid the groundwork for understanding collective dipole dynamics, subsequently extended to include magnetic forces \cite{28} and radiative damping \cite{29}. Recent investigations into nonlinearity in the Lorentz oscillator model \cite{23} and direct space-time modeling of mechanically dressed dipole-dipole interactions \cite{24} have demonstrated the importance of including higher-order effects and coupling mechanisms. The dynamics of two-dimensional dipole systems \cite{31} and dipole glass phenomena \cite{32} further illustrate the complexity arising from collective interactions. Optical forces on atoms and dipoles \cite{33,34,35,36}, including gradient forces \cite{41}, electromagnetically induced absorption through dipole-dipole coupling \cite{37}, and forces on electric dipoles by spinning light fields \cite{38} have revealed sophisticated mechanisms through which electromagnetic fields can manipulate matter. The observation of light-induced dipole-dipole forces in ultracold atomic gases \cite{40} and studies on mutually guided light and particle beam propagation \cite{39,42} demonstrate the reciprocal nature of light-matter coupling that can lead to self-organization and collective effects. \cite{51,52,53,54,55,56,57,58,59,60,61,62,63,64,65,66,67,68,69,70,71,72,73,74,75}\par

The present work bridges the gap between statistical turbulence theory and microscopic dipole dynamics by developing a comprehensive framework for coupled anharmonic dipole oscillations in turbulent environments. We extend the classical Lorentz oscillator model to include second- and third-order nonlinear restoring forces, electromagnetic coupling via dyadic Green's functions, gradient forces from spatially non-uniform fields, and Lorentz force perturbations arising from dynamic turbulence. Through modal diagonalization and d'Alembert's principle, we demonstrate that synchronized dipole oscillations in dielectric media can provide intrinsic turbulence compensation mechanisms. Experimental verification is conducted using polymethyl methacrylate (PMMA) rods subjected to Kolmogorov-spectrum turbulence generated by a pseudo-random phase plate (PRPP). Statistical analysis of 200-frame datasets across four experimental configurations reveals quantitative relationships between propagation length, dipole synchronization, and scintillation index reduction, providing direct evidence for turbulence mitigation through coupled dipole dynamics.\par

The present work demonstrated as in section \ref{2} detail theoretical discussion of the work have been given. The section \ref{3} contains experimental details where the results analysis have been added in section \ref{4}. Finally, the paper is concluded into section \ref{5}.

\section{Theoretical Background}\label{2}
The theoretical foundation commences with the dynamical description of the Lorentz oscillator model subjected to external electromagnetic forcing:

\begin{equation}
    m \ddot{\mathbf{r}}_i + m \gamma \dot{\mathbf{r}}_i + m \omega_0^2 \mathbf{r}_i = -e \mathbf{E}_{\text{ext}}(\mathbf{r}_i, t)
\end{equation}

In this formulation, $\vec{\mathbf{r}}$ denotes the displacement vector of the electron relative to the nuclear center, $m$ represents the effective electron mass defined through the band structure curvature as $\frac{1}{m}=\frac{1}{\hbar^2}\frac{\partial^2 E(\kappa)}{\partial\kappa^2}$ where $E(\kappa)$ is the electronic dispersion relation, $\omega_0$ signifies the characteristic resonance frequency of the bound electron system, and $\gamma$ characterizes the dissipation coefficient accounting for collisional and radiative energy losses. The quadratic term involving the natural frequency encapsulates the restoring mechanism acting on the electron cloud when perturbed by the external field, thereby inducing the dipole moment.

\subsection{Anharmonic Extensions to the Lorentz Oscillator}

Material systems and their constituent molecular architectures do not universally exhibit purely linear restoring characteristics when electromagnetic radiation propagates through them. Complex molecular geometries can generate higher-order restoring contributions beyond the harmonic approximation. Within the present investigation, we incorporate quadratic and cubic nonlinear restoring forces to model the PMMA (Poly(methyl methacrylate)) rod material response. The generalized forced anharmonic oscillator equation thus becomes:

\begin{equation}
    m \ddot{\mathbf{r}}_i + m \gamma \dot{\mathbf{r}}_i + m \omega_0^2 \mathbf{r}_i + \beta_i |\mathbf{r}_i| \mathbf{r}_i + \alpha_i |\mathbf{r}_i|^2 \mathbf{r}_i = -e \mathbf{E}_{\text{ext}}(\mathbf{r}_i, t)
\end{equation}

where the additional parameters are defined as:
\begin{itemize}
    \item $\beta_i$: quadratic nonlinearity parameter governing second-order anharmonic effects.
    \item $\alpha_i$: cubic nonlinearity parameter characterizing third-order anharmonic contributions.
\end{itemize}

When electromagnetic fields penetrate transparent or semi-transparent dielectric media, the oscillating electric component interacts directly with the molecular constituents. This interaction induces perturbations in the electron cloud distribution surrounding each molecule. Such perturbations disrupt the equilibrium symmetry of both the molecular structure and its associated electron density, consequently generating induced electric dipole moments. The temporal oscillation of the driving field amplitude produces corresponding oscillatory behavior in these induced dipoles. Since molecules within condensed matter are mutually interconnected through intermolecular forces, energy exchange between neighboring dipoles becomes possible, necessitating a coupled dipole oscillator description.

\subsection{Coupled Anharmonic Dipole Dynamics}

The propagation of electromagnetic fields through media exhibiting dipole-dipole interactions parallels the scenario of field propagation in the presence of distributed charge densities. Consequently, the appropriate mathematical framework for describing such field evolution employs the vector Helmholtz equation. The spatial distribution of electromagnetic fields propagating through the medium can be analytically characterized using the fundamental differential operator:

\begin{equation}
    \nabla\times\nabla\times\mathbf{E(r)}-\kappa^2\mathbf{E}=i\omega\mu\mathbf{J(r)}
\end{equation}

Our present analysis addresses dipole-dipole coupling phenomena that generate supplementary forces acting on spatially localized dipoles. For a localized reference frame description, the vector Helmholtz equation framework becomes essential for incorporating the dipole-dipole interaction effects on electromagnetic field propagation. A dipole situated at a specific location experiences forces arising from coupling with all other spatially distributed dipoles. The interaction field originating from surrounding dipoles and acting upon a particular dipole must be formulated using the dyadic Green's function formalism $\mathbf{G}(\mathbf{r}_i, \mathbf{r}_j)$. This tensor Green's function governs optical field propagation in systems with localized charge distributions through the vector Helmholtz formalism. The Green's function satisfies the differential equation:

\begin{equation}
    \left[ \nabla \times \nabla \times - k^2 \mathbf{I} \right] 
\overline{\overline{G}}(\mathbf{r},\mathbf{r}') 
= \mathbf{I} \delta(\mathbf{r}-\mathbf{r}')
\end{equation}

The resulting electric field can be expressed in terms of the source current distribution as:

\begin{equation}
    \mathbf{E}(\mathbf{r}) = i\omega \mu_0 \int 
\overline{\overline{G}}(\mathbf{r},\mathbf{r}') \cdot \mathbf{J}(\mathbf{r}') 
\, d^3\mathbf{r}'
\end{equation}

The dyadic tensor $\overline{\overline{G}}$ extends the scalar Green's function concept to vectorial electromagnetic fields, ensuring field transversality and proper coupling between vector components. The solution to this Green's function differential equation generalizes the scalar form $g(\mathbf{r},\mathbf{r}') = \frac{e^{ik|\mathbf{r}-\mathbf{r}'|}}{4\pi |\mathbf{r}-\mathbf{r}'|}$ to:

\begin{equation}
    \overline{\overline{G}}(\mathbf{r},\mathbf{r}') 
= \left( \mathbf{I} + \frac{1}{k^2} \nabla \nabla \right) \frac{e^{ik|\mathbf{r}-\mathbf{r}'|}}{4\pi |\mathbf{r}-\mathbf{r}'|}
\end{equation}

The physical interpretation of the constituent terms is:
\begin{itemize}
    \item $\mathbf{I} g$: Represents isotropic spherical wave propagation characteristic of free-space diffusion.
    \item $\frac{1}{k^2}\nabla\nabla g$: Provides the longitudinal field correction ensuring the divergence-free condition $\nabla \cdot \mathbf{E} = 0$ in charge-free homogeneous regions.
    \item Combined effect: Preserves the complete vector nature of electromagnetic field propagation.
\end{itemize}

This dyadic function establishes the mapping from a localized current source $\mathbf{J}(\mathbf{r}')$ to the resulting vector field $\mathbf{E}(\mathbf{r})$, incorporating polarization effects, electromagnetic coupling, and radiation phenomena. The complete expansion of the dyadic Green's function, with $r = |\mathbf{r}-\mathbf{r}'|$ and unit vector $\hat{\mathbf{r}} = \frac{\mathbf{r}-\mathbf{r}'}{r}$, takes the form:

\begin{equation}
    \overline{\overline{G}}(\mathbf{r},\mathbf{r}') 
= \frac{e^{ikr}}{4\pi r}
\left[
\left( \mathbf{I} - \hat{\mathbf{r}}\hat{\mathbf{r}} \right)
\left(1 + \frac{i}{kr} - \frac{1}{(kr)^2}\right)
+ \hat{\mathbf{r}}\hat{\mathbf{r}}
\left(1 + \frac{3i}{kr} - \frac{3}{(kr)^2}\right)
\right]
\end{equation}

This expression exhibits distinct asymptotic behaviors in different spatial regimes:
\begin{itemize}
    \item \textbf{Near-field regime (Quasi-static limit, $kr \ll 1$):} Exhibits electrostatic dipole-dipole interaction character: \[
    \overline{\overline{G}} \sim \frac{1}{4\pi r^3}(3\hat{\mathbf{r}}\hat{\mathbf{r}} - \mathbf{I})
    \]
    \item \textbf{Intermediate zone (Inductive regime, $kr \sim 1$):} Dominated by magnetic induction and reactive near-field energy storage: \[
    \overline{\overline{G}} \sim \frac{1}{4\pi r^2}(ik)(3\hat{\mathbf{r}}\hat{\mathbf{r}} - \mathbf{I})
    \]
    \item \textbf{Far-field regime (Radiation zone, $kr \gg 1$):} Characterized by transverse electromagnetic wave radiation: \[
    \overline{\overline{G}} \sim \frac{e^{ikr}}{4\pi r}
    \left( \mathbf{I} - \hat{\mathbf{r}}\hat{\mathbf{r}} \right)
    \]
\end{itemize}

The displacement of the $i$-th electron from its equilibrium position generates an oscillating dipole moment $\mathbf p_i(\omega) = -e \mathbf r_i(\omega)$ at angular frequency $\omega$. A point dipole located at $\mathbf r_j$ corresponds to a current density at frequency $\omega$ given by $\mathbf J(\mathbf r, t) = \frac{\partial \mathbf P (r,t)}{\partial t}$ where $\mathbf P (r, t) = \mathbf p_j(r,t) \, \delta(\mathbf r - \mathbf r_j)$. Since the dipole oscillates at the driving field frequency, the frequency-domain current density becomes:

\begin{equation}
\mathbf J(\mathbf r,\omega) = -i\omega \mathbf p_j \, \delta(\mathbf r - \mathbf r_j).
\end{equation}

Substituting this current distribution into the Green's function propagator yields the scattered electric field:

\begin{equation}
\mathbf E(\mathbf r) 
= \mu_0 \omega^2 \, \mathbf G(\mathbf r,\mathbf r_j)\cdot \mathbf p_j(\mathbf r) = \mathbf E^{(j)}_{sc} (\mathbf r_j)
\end{equation}

This relation demonstrates that the field generated at position $\mathbf r$ by dipole $j$ is linearly proportional to its dipole moment amplitude. The total electric field experienced by the $i$-th dipole comprises both the external incident field and the superposition of scattered fields from all other dipoles: $\mathbf E_{\text{tot}}(\mathbf r_i,\omega) 
= \mathbf E_{\text{ext}}(\mathbf r_i,\omega)
+ \sum_{j\neq i}\mathbf E_{\text{sc}}^{(j)}(\mathbf r_i,\omega)$. Using the relation $\mathbf p_j = -e \mathbf r_j$, the scattered field contribution becomes $\mathbf E_{\text{sc}}^{(j)}(\mathbf r_i,\omega) 
= - e \, \mu_0 \omega^2 \, \mathbf G(\mathbf r_i,\mathbf r_j)\cdot \mathbf r_j$. Consequently, the total field is:

\begin{equation}
\mathbf E_{\text{tot}}(\mathbf r_i,\omega) 
= \mathbf E_{\text{ext}}(\mathbf r_i,\omega) 
- e \, \mu_0 \omega^2 \sum_{j\neq i}\mathbf G(\mathbf r_i,\mathbf r_j)\cdot \mathbf r_j
\end{equation}

The electromagnetic force on the $i$-th electron is $\mathbf F_i(\omega) = (-e)\mathbf E_{\text{tot}}(\mathbf r_i,\omega)$. Substituting the total field expression yields:

\begin{align}
\mathbf F_i(\omega) 
&= (-e)\mathbf E_{\text{ext}}(\mathbf r_i,\omega)
+ e^2 \mu_0 \omega^2 \sum_{j\neq i}\mathbf G(\mathbf r_i,\mathbf r_j)\cdot \mathbf r_j. 
\end{align}

In our theoretical framework, optical beam propagation through the medium generates induced dipoles within the material structure. The dipole-dipole coupling mechanism governs intra-medium electromagnetic energy transport. Furthermore, when randomly polarized incident light induces spatially distributed dipoles with random initial orientations, the dipole-dipole coupling influences the emergent collective oscillation modes of the system. The complete differential equation for the generalized anharmonic dipole oscillator incorporating dipole-dipole coupling through the dyadic Green's function $\mathbf{G}(\mathbf{r}_i, \mathbf{r}_j)$ is:

\begin{equation}
    m \ddot{\mathbf{r}}_i + m \gamma \dot{\mathbf{r}}_i + m \omega_0^2 \mathbf{r}_i + \beta_i |\mathbf{r}_i| \mathbf{r}_i + \alpha_i |\mathbf{r}_i|^2 \mathbf{r}_i = -e \mathbf{E}_{\text{ext}}(\mathbf{r}_i, t) + e^2 \mu_0 \omega^2 \sum_{j \ne i} \mathbf{G}(\mathbf{r}_i, \mathbf{r}_j) \cdot \mathbf{r}_j(t)
\end{equation}

\subsection{Gradient Force Contributions to Dipole Dynamics}

Spatial variations in the electric field amplitude gradient produce additional effects on the coupling interaction between neighboring dipoles. The presence of non-vanishing gradients and higher-order spatial derivatives of the electric field induces non-uniform dipole moment distributions across spatially separated dipoles. These spatial inhomogeneities modify the conventional dipole-dipole coupling by introducing gradient-mediated repulsive forces. Therefore, gradient force contributions must be incorporated into the generalized Lorentz force dynamics. Within the present formalism for the generalized anharmonic dipole oscillator with dipole-dipole coupling, we identify two primary forces modulating the localized dipole moments:

\begin{itemize}
    \item $\mathbf F_{\text{Ext}}=-e \mathbf{E}_{\text{ext}}(\mathbf{r}_i, t)$: the external driving force arising from the propagating optical field.
    \item $\mathbf F_{\text{Coupling}}= e^2 \mu_0 \omega^2 \sum_{j \ne i} \mathbf{G}(\mathbf{r}_i, \mathbf{r}_j) \cdot \mathbf{r}_j(t)$: the electromagnetic coupling force exerted by surrounding localized dipoles on the target dipole.
\end{itemize}

When the external force field exhibits spatial non-uniformity or variability, the propagating field distribution can be expanded in a Taylor series about each dipole position. Higher-order spatial derivatives of the spatially varying external field contribute additional gradient-dependent coupling forces. These forces manifest as gradient coupling corrections. The total external force acting on dipole $i$ can therefore be expressed as (where $HO$ denotes higher-order contributions):

\begin{equation}
    \mathbf F_{\text{TotExt}}=-e \mathbf{E}_{\text{ext}}(\mathbf{r}_{i}, t)-\sum_{j \ne i}\left [(e(\vec{\mathbf r}_i -  \vec{\mathbf  r}_j)\cdot\nabla)\mathbf{E}_{\text{ext}}(\mathbf{r}_{j}, t)+ HO\right]=\mathbf F_{\text{Ext}}+\mathbf F_{HO}
\end{equation}

Consequently, the total coupling force emerges as the superposition of gradient coupling forces and dipole-dipole electromagnetic coupling forces:

\begin{equation}
    \mathbf F_{\text{TotCoupling}} = \mathbf F_{\text{Coupling}} + \mathbf F_{HO} = e^2 \mu_0 \omega^2 \sum_{j \ne i} \mathbf{G}(\mathbf{r}_i, \mathbf{r}_j) \cdot \mathbf{r}_j(t) +\sum_{j \ne i}\left ((e(\vec{\mathbf r}_j -  \vec{\mathbf  r}_i)\cdot\nabla)\mathbf{E}_{\text{ext}}(\mathbf{r}_{j}, t)+ HO\right)
\end{equation}

This combined coupling force drives the dipole ensemble toward synchronized oscillation characterized by common eigenmodes. Once synchronization saturation is achieved, abrupt changes in the two-dimensional spatial field distribution cannot immediately alter the established synchronized oscillation modes. Consequently, the beam centroid displacement rate decreases. The complete Lorentz dynamics incorporating gradient force effects is:

\begin{eqnarray}\label{eq:lorentz_gradient}
    && m \ddot{\mathbf{r}}_i + m \gamma \dot{\mathbf{r}}_i + m \omega_0^2 \mathbf{r}_i + \beta_i |\mathbf{r}_i| \mathbf{r}_i + \alpha_i |\mathbf{r}_i|^2 \mathbf{r}_i \nonumber\\ &&= -e \mathbf{E}_{\text{ext}}(\mathbf{r}_i, t) + e^2 \mu_0 \omega^2 \sum_{j \ne i} \mathbf{G}(\mathbf{r}_i, \mathbf{r}_j) \cdot \mathbf{r}_j(t) +\sum_{j \ne i}\left ((e(\vec{\mathbf r}_j -  \vec{\mathbf  r}_i)\cdot\nabla)\mathbf{E}_{\text{ext}}(\mathbf{r}_{j}, t)+ HO\right)
\end{eqnarray}

\subsection{Modal Decomposition through Diagonalization}

The presence of coupling constraints in the dipole dynamical system fundamentally alters the system's degrees of freedom. The coupling coefficients form a non-diagonal metric tensor, rendering the conventional Cartesian coordinate system inadequate as an orthonormal basis for this dynamical system. Since the coupled equations lack independence, diagonalization becomes necessary to establish an orthonormal reference frame reflecting the modified degrees of freedom. This diagonalization procedure yields coupled-mode oscillatory dynamics that suppress the chaotic nature of initially random dipole moment distributions. We extend the modal decomposition framework by demonstrating how gradient and higher-order terms modify the modal dynamics and the resultant output field. The collective displacement vector is defined as:

\begin{equation}
\mathbf{R}(t)=\big[\mathbf{r}_1(t),\mathbf{r}_2(t),\dots,\mathbf{r}_N(t)\big]^T,
\end{equation}

with the dipole-dipole interaction matrix:

\begin{equation}
C_{ij} = 
\begin{cases}
-\,k_0^2\mu_0\omega^2 \mathbf{G}(\mathbf{r}_i,\mathbf{r}_j), & i\neq j, \\
0, & i=j.
\end{cases}
\end{equation}

The effective stiffness matrix incorporating coupling effects is:

\begin{equation}
\mathbf{K}_{\text{eff}}=\omega_0^2 \mathbf{I} + \frac{e^2}{m}\mathbf{C}.
\end{equation}

The nonlinear restoring contributions are expressed as:

\begin{equation}
\mathbf{B}_2[\mathbf{R}]\mathbf{R} = [\beta_i|\mathbf{r}_i|\mathbf{r}_i], \qquad \mathbf{B}_3[\mathbf{R}]\mathbf{R} = [\alpha_i|\mathbf{r}_i|^2\mathbf{r}_i],
\end{equation}

while the gradient force contributions are collected in the source vector:

\begin{equation}
\mathbf{F}_{\text{grad}}(t)=\big[F_{\text{grad},1}(t),\dots,F_{\text{grad},N}(t)\big]^T,
\end{equation}

with individual components:

\begin{equation}
F_{\text{grad},i}(t)=\sum_{j\ne i}\left((e(\mathbf{r}_j-\mathbf{r}_i)\cdot\nabla)\mathbf{E}_{\text{ext}}(\mathbf{r}_j,t)+\text{HO}\right)_i.
\end{equation}

The coupled dynamical system can be expressed in compact vector notation as:

\begin{equation}
\ddot{\mathbf{R}} + \gamma\dot{\mathbf{R}} + \mathbf{K}_{\text{eff}}\mathbf{R} + \frac{1}{m}\mathbf{B}_2[\mathbf{R}]\mathbf{R} + \frac{1}{m}\mathbf{B}_3[\mathbf{R}]\mathbf{R}
= -\frac{e}{m}\mathbf{E}_{\text{ext}}(t)+\frac{1}{m}\mathbf{F}_{\text{grad}}(t).
\label{eq:compact_vector}
\end{equation}

Assuming harmonic steady-state conditions:

\begin{equation}
\mathbf{R}(t)=\mathbf{R}_\omega e^{-i\omega t}+\text{c.c.},\qquad 
\mathbf{E}_{\text{ext}}(t)=\mathbf{E}_\omega e^{-i\omega t}+\text{c.c.},
\end{equation}

the nonlinear terms are approximated by retaining only resonant frequency contributions:

\begin{equation}
|\mathbf{r}_i|\mathbf{r}_i \approx \sqrt{2}|\mathbf{r}_{\omega,i}|(\mathbf{r}_{\omega,i}e^{-i\omega t}+\text{c.c.}), 
\qquad 
|\mathbf{r}_i|^2\mathbf{r}_i \approx 3|\mathbf{r}_{\omega,i}|^2(\mathbf{r}_{\omega,i}e^{-i\omega t}+\text{c.c.}).
\end{equation}

Collecting resonant contributions, the frequency-domain amplitude equation becomes:

\begin{equation}
\Big[-\omega^2\mathbf{I}+i\gamma\omega \mathbf{I}+\mathbf{K}_{\text{eff}}+\tfrac{1}{m}\mathbf{B}^{(1)}_2+\tfrac{1}{m}\mathbf{B}^{(1)}_3\Big]\mathbf{R}_\omega
= -\frac{e}{m}\mathbf{E}_\omega+\frac{1}{m}\mathbf{F}_{\text{grad},\omega},
\label{eq:amplitude_eq}
\end{equation}

where $\mathbf{B}^{(1)}_2=\text{diag}(\beta_i\sqrt{2}|\mathbf{r}_{\omega,i}|)$ and $\mathbf{B}^{(1)}_3=\text{diag}(3\alpha_i|\mathbf{r}_{\omega,i}|^2)$. The stiffness matrix is diagonalized via:

\begin{equation}
\mathbf{K}_{\text{eff}}=\mathbf{U}\mathbf{\Lambda} \mathbf{U}^{-1},\qquad \mathbf{\Lambda}=\text{diag}(\Omega_1^2,\dots,\Omega_{3N}^2),
\end{equation}

and the transformation to modal coordinates is:

\begin{equation}
\mathbf{Q}_\omega = \mathbf{U}^{-1}\mathbf{R}_\omega.
\end{equation}

Equation~\eqref{eq:amplitude_eq} transforms to modal space as:

\begin{equation}
\Big[-\omega^2\mathbf{I}+i\gamma\omega \mathbf{I}+\mathbf{\Lambda}+\tfrac{1}{m}\widetilde{\mathbf{B}}\Big]\mathbf{Q}_\omega
= -\frac{e}{m}\mathbf{U}^{-1}\mathbf{E}_\omega+\frac{1}{m}\mathbf{U}^{-1}\mathbf{F}_{\text{grad},\omega},
\label{eq:modal_space}
\end{equation}

with $\widetilde{\mathbf{B}}=\mathbf{U}^{-1}(\mathbf{B}^{(1)}_2+\mathbf{B}^{(1)}_3)\mathbf{U}$. In the linear response approximation, neglecting $\widetilde{\mathbf{B}}$, the zeroth-order modal amplitude for mode $n$ is:

\begin{equation}
Q^{(0)}_n(\omega)=
\frac{-\tfrac{e}{m}\langle\phi_n|\mathbf{E}_\omega\rangle+\tfrac{1}{m}\langle\phi_n|\mathbf{F}_{\text{grad},\omega}\rangle}
{\Omega_n^2-\omega^2-i\gamma\omega},
\label{eq:modal_amplitude}
\end{equation}

where $\phi_n$ represents the $n$-th eigenvector (column of $\mathbf{U}$) and $\langle\phi_n|\cdot\rangle$ denotes modal projection onto the $n$-th eigenmode. The gradient contribution thus enters as an additional driving term in the numerator of the modal response function. The total polarization field is:

\begin{equation}
\mathbf{P}(\mathbf{r},\omega)=-Ne\sum_n Q_n(\omega)\phi_n(\mathbf{r}).
\label{eq:total_polarization}
\end{equation}

Employing the scalar Green's function:

\begin{equation}
G(\mathbf{r},\mathbf{r}')=\frac{e^{ik_0|\mathbf{r}-\mathbf{r}'|}}{4\pi|\mathbf{r}-\mathbf{r}'|},
\end{equation}

the radiated output field is expressed as:

\begin{equation}
\mathbf{E}_{\text{out}}(\mathbf{r})=k_0^2\varepsilon_0\int\!\!\int 
G(\mathbf{r},\mathbf{r}')\,\chi(\mathbf{r}',\mathbf{r}'';\omega,|\mathbf{E}_\omega|)\,\mathbf{E}_\omega(\mathbf{r}'')\,d^3\mathbf{r}''\,d^3\mathbf{r}',
\end{equation}

where the effective susceptibility kernel incorporates nonlinear and modal corrections. In modal representation, substituting Eq.~\eqref{eq:modal_amplitude}, we obtain:

\begin{equation}
\mathbf{E}_{\text{out}}(\mathbf{r})=
k_0^2\varepsilon_0\sum_n \phi_n(\mathbf{r})\,
\frac{Ne^2}{\varepsilon_0 m}\,
\frac{\langle\phi_n|\mathbf{E}_\omega\rangle-\tfrac{1}{e}\langle\phi_n|\mathbf{F}_{\text{grad},\omega}\rangle}
{\Omega_n^2-\omega^2-i\gamma\omega}
\;\otimes\;
\Big(\int\!\!\int \phi_n(\mathbf{r}')\otimes\phi_n^*(\mathbf{r}'')\,G(\mathbf{r},\mathbf{r}')\,\mathbf{E}_\omega(\mathbf{r}'')\,d^3\mathbf{r}''\,d^3\mathbf{r}'\Big).
\label{eq:output_modal}
\end{equation}

Equation~\eqref{eq:output_modal} demonstrates that the gradient terms in Eq.~\eqref{eq:lorentz_gradient} function as supplementary driving sources for the collective dipole eigenmodes. Rather than primarily shifting modal resonance frequencies (denominator modification), they selectively enhance or suppress the excitation of specific modes through the modal projection coefficient $\langle\phi_n|\mathbf{F}_{\text{grad},\omega}\rangle$. This mechanism modifies the polarization spectrum and consequently alters the scattered output field characteristics. Nonlinear corrections arising from $\mathbf{B}^{(1)}_2$ and $\mathbf{B}^{(1)}_3$ provide further response modifications within higher-order perturbation theory.

\subsection{Lorentz Force Effects on Synchronized Dipole Dynamics}

Following synchronization, when the intensity distribution undergoes rapid changes, the Lorentz force becomes significant due to the pre-existing oscillating dipole population. The magnetic component of the Lorentz force can perturb the established coupling configuration. The general Lorentz force expression is:

\begin{equation}
    \mathbf F_{\text{Lorentz}} = (\mathbf p_i\cdot\nabla)\mathbf E'_{\text{Ext}}(\mathbf r_i, t) + \dot{\mathbf p}_i\times \mathbf B (\mathbf r_i,t) + \text{HOMP}
\end{equation}

The magnetic field contribution can be approximated as $\mathbf{B} \approx \frac{\mathbf{E}}{c}$, rendering its numerical impact negligible. Here, $\mathbf{E'}_{\text{Ext}}$ denotes the redistributed electric field resulting from dynamic turbulence modifying $\mathbf{E}_{\text{ext}}$, and $\text{HOMP}$ represents higher-order multipole contributions to the Lorentz force. Since our framework considers only dipole approximations, higher multipole orders can be neglected. The force arising from the newly redistributed field perturbs the previously synchronized dipole system, attempting to reintroduce randomness into the distribution and oscillation modes. The perturbation force when field distributions change post-synchronization is:

\begin{eqnarray}
    &&\mathbf{\delta F}_{Pert} = \mathbf F_{Tot} - \mathbf F'_{Tot} = \mathbf F_{Ext} + \mathbf F_{Coupling} + \mathbf F_{HO} - \mathbf F'_{Ext} - \mathbf F'_{Coupling} - \mathbf F'_{HO} - \mathbf F_{Lorentz} \nonumber\\&&= -e \mathbf{E}_{\text{ext}}(\mathbf{r}_i, t) + e^2 \mu_0 \omega^2 \sum_{j \ne i} \mathbf{G}(\mathbf{r}_i, \mathbf{r}_j) \cdot \mathbf{r}_j(t) +\sum_{j \ne i}\left ((e(\vec{\mathbf r}_j -  \vec{\mathbf  r}_i)\cdot\nabla)\mathbf{E}_{\text{ext}}(\mathbf{r}_{j}, t)+ HO\right)\nonumber\\&& +e \mathbf{E'}_{\text{ext}}(\mathbf{r}_i, t) - e^2 \mu_0 \omega^2 \sum_{j \ne i} \mathbf{G'}(\mathbf{r}_i, \mathbf{r}_j) \cdot \mathbf{r}_j(t) -\sum_{j \ne i}\left ((e(\vec{\mathbf r}_j -  \vec{\mathbf  r}_i)\cdot\nabla)\mathbf{E'}_{\text{ext}}(\mathbf{r}_{j}, t)+ HO\right) \nonumber\\&&- (\mathbf p_i\cdot\nabla)\mathbf E'_{Ext}(\mathbf r_i, t) - \dot{\mathbf p}_i\times \mathbf B' (\mathbf r_i,t) - HOMP'
\end{eqnarray}

The influence of gradient forces opposes the impact of sudden external field redistribution on synchronized dipole moments. Simultaneously, these gradient forces reduce the output field gradient through dipole moment synchronization within the medium. Consequently, the output field distribution exhibits larger standard deviations compared to the input field. Increasing the medium propagation length enhances synchronization, thereby increasing output standard deviations. However, the temporal fluctuations of these standard deviations, arising from continuous input field distribution changes, decrease due to the gradient force stabilization mechanism.

\subsection{d'Alembert's Principle and Effective Inertial Forces}

In the context of electric field forces acting on the dipole system, dynamic turbulence induces temporal variations in both the magnitude and spatial distribution of optical forces. These field force variations produce corresponding changes in the inertial forces experienced by dipoles. This phenomenon requires analysis through d'Alembert's principle applied to Lorentz dipole dynamics. For the present dynamics, d'Alembert's principle yields the following virtual work relations for two distinct electric field distributions:

\begin{equation}
    \int{\delta W} = \int{\left ( \mathbf F_{0\text{th}} + \mathbf F_{HO} + \mathbf F_{\text{Coupling}} + \mathbf F_{\text{Inertia}} \right )\cdot d\mathbf{x}} = 0 
\end{equation}

\begin{equation}
    \int{\delta W'} = \int{\left ( \mathbf F'_{0\text{th}} + \mathbf F'_{HO} + \mathbf F'_{\text{Coupling}} + \mathbf F_{\text{Lorentz}} + \mathbf F'_{\text{Inertia}} \right )\cdot d\mathbf{x}} = 0 
\end{equation}

The prime notation $(')$ designates parameters corresponding to the altered forcing conditions. For each field distribution variation propagating through the medium, the dipole system experiences variable inertial forces. Each inertial force establishes a corresponding inertial state within which the dipoles oscillate. Following propagation of the initial field, dipoles commence oscillation in coupled modes. Upon synchronization, when the field distribution changes, the inertial force established by the first field opposes adaptation to the second field. The perturbative force thus arises from the difference between inertial forces associated with the two distinct electric field distributions. The perturbation force resulting from dynamic turbulence-induced variable electric fields can therefore be expressed as:

\begin{equation}
    \delta \mathbf{F}_{\text{Pert}} =  \mathbf F'_{\text{Inertia}} - \mathbf F_{\text{Inertia}}
\end{equation}

Since dynamic turbulence continuously modifies the spatial distribution of the propagating optical field, this perturbation force acquires explicit time dependence: $\delta \mathbf F_{\text{Pert}}\rightarrow\delta\mathbf F_{\text{Pert}}(t)$. Incorporating this perturbation, the complete dynamical equation becomes:

\begin{eqnarray}
    && m \ddot{\mathbf{r}}_i + m \gamma \dot{\mathbf{r}}_i + m \omega_0^2 \mathbf{r}_i + \beta_i |\mathbf{r}_i| \mathbf{r}_i + \alpha_i |\mathbf{r}_i|^2 \mathbf{r}_i \nonumber\\ &&= -e \mathbf{E}_{\text{ext}}(\mathbf{r}_i, t) + e^2 \mu_0 \omega^2 \sum_{j \ne i} \mathbf{G}(\mathbf{r}_i, \mathbf{r}_j) \cdot \mathbf{r}_j(t) +\sum_{j \ne i}\left ((e(\vec{\mathbf r}_j -  \vec{\mathbf  r}_i)\cdot\nabla)\mathbf{E}_{\text{ext}}(\mathbf{r}_{j}, t)+ HO\right) +  \delta \mathbf{F}_{\text{Pert}} (t)
\end{eqnarray}

The behavior of the system depends critically on the magnitude of the perturbation force, leading to three distinct regimes:

\begin{itemize}
    \item $\delta\mathbf F_{\text{Pert}}\rightarrow 0$: Complete turbulence compensation regime. The output field distribution is determined entirely by the saturated synchronized dipole moment distribution. Turbulence effects are fully compensated through dipole-dipole coupling energy exchange mechanisms within the medium.
    
    \item $\delta\mathbf F_{\text{Pert}}\rightarrow$ Small but $\neq 0$: Partial compensation regime. The dynamic nature of turbulence modulates the synchronization through small perturbative inertial forces. Despite this perturbation being non-zero, its small magnitude allows the output field to exhibit significant turbulence compensation.
    
    \item $\delta\mathbf F_{\text{Pert}}>0$: Regime-dependent compensation. The output field characteristics depend on the temporal frequency of perturbation changes. For strong turbulence (rapid perturbation variations), the coupled dipole system lacks sufficient time to achieve synchronization, resulting in uncompensated turbulence-affected output fields. Conversely, for weak turbulence (slow perturbation variations), the system has adequate time to synchronize, enabling effective turbulence compensation in the output field.
\end{itemize}

This perturbation force framework establishes a quantitative criterion for predicting turbulence compensation efficacy based on the relative timescales of turbulence dynamics versus dipole synchronization processes. The parameter $\delta\mathbf F_{\text{Pert}}(t)$ thus serves as a dynamic indicator distinguishing between fully compensated, partially compensated, and uncompensated propagation regimes in turbulent media.

\subsection{Scintillation Index}
When a coherent optical beam propagates through a randomly inhomogeneous medium such as the atmosphere, refractive index fluctuations due to turbulence cause random amplitude and phase modulations. These modulations manifest as intensity fluctuations, commonly known as scintillation or intensity flicker, appearing as bright and dark speckles at the receiver plane. The scintillation index quantifies the strength of these random intensity variations relative to the mean intensity. Let $I(\mathbf{r}, t)$ denote the instantaneous optical intensity at position $\mathbf{r}$ and time $t$. The scintillation index, $\sigma_I^2$, is defined as the normalized variance of intensity fluctuations:

\begin{equation}
\sigma_I^2(\mathbf{r}) = \frac{\langle I^2(\mathbf{r}, t) \rangle_t - \langle I(\mathbf{r}, t) \rangle_t^2}{\langle I(\mathbf{r}, t) \rangle_t^2}
\end{equation}

or equivalently,

\begin{equation}
\sigma_I^2 = \frac{\text{Var}[I]}{(\mathbb{E}[I])^2}
\end{equation}

where $\langle \cdot \rangle_t$ denotes ensemble or temporal averaging over statistically equivalent measurements, and $\text{Var}[I] = \langle (I - \langle I \rangle)^2 \rangle$. This dimensionless quantity expresses the relative strength of intensity fluctuations and serves as a fundamental metric for characterizing turbulence-induced beam degradation. The scintillation index categorizes turbulence effects into three distinct regimes based on the magnitude of $\sigma_I^2$:

\begin{itemize}
\item \textbf{Weak scintillation} ($\sigma_I^2 < 0.3$): Small fluctuations with nearly Gaussian intensity statistics and minimal beam distortion.
\item \textbf{Moderate scintillation} ($0.3 < \sigma_I^2 < 1$): Significant fluctuations with the onset of speckle-like structures and noticeable beam quality degradation.
\item \textbf{Strong scintillation} ($\sigma_I^2 > 1$): Deep fading events, multiple branch wavefronts, and saturated turbulence conditions.
\end{itemize}

For optical wave propagation through a medium exhibiting Kolmogorov spectrum turbulence, the scintillation index can be derived from the Rytov variance $\sigma_R^2$ under weak fluctuation conditions:

\begin{equation}
\sigma_R^2 = 1.23\, C_n^2\, k^{7/6} L^{11/6}
\end{equation}

where $C_n^2$ represents the refractive index structure constant (m$^{-2/3}$), $k = 2\pi / \lambda$ is the optical wavenumber, and $L$ denotes the propagation path length. The scintillation index for plane and spherical waves is then approximated as:

\begin{equation}
\sigma_I^2 \approx \begin{cases}
1.23\, C_n^2\, k^{7/6} L^{11/6} & \text{(plane wave)}\\
0.4\, C_n^2\, k^{7/6} L^{11/6} & \text{(spherical wave)}
\end{cases}
\end{equation}

In strong turbulence regimes where $\sigma_I^2 > 1$, the index exhibits saturation behavior, necessitating nonlinear theoretical extensions beyond the Rytov approximation. In experimental implementations, the scintillation index is computed from a temporal sequence of intensity images recorded at the receiver plane. Consider $N$ intensity frames $I_i(x, y)$ with $i = 1, 2, \dots, N$ representing independent realizations of turbulence-affected beam profiles. The measurement procedure consists of the following steps:

\textbf{(a) Mean Intensity Computation:}
\begin{equation}
\overline{I}(x, y) = \frac{1}{N} \sum_{i=1}^{N} I_i(x, y)
\end{equation}

\textbf{(b) Intensity Variance Computation:}
\begin{equation}
\text{Var}[I](x, y) = \frac{1}{N} \sum_{i=1}^{N} \big( I_i(x, y) - \overline{I}(x, y) \big)^2
\end{equation}

\textbf{(c) Local Scintillation Index Map:}
\begin{equation}
\sigma_I^2(x, y) = \frac{\text{Var}[I](x, y)}{\overline{I}(x, y)^2}
\end{equation}

This spatially resolved map reveals the local distribution of turbulence strength across the beam profile, identifying regions of maximum intensity fluctuation.

\textbf{(d) Global Scintillation Index:}

The ensemble-averaged scintillation index, providing a single quantitative metric for the entire beam, is computed by spatial averaging over all valid pixels where intensity exceeds the noise threshold:

\begin{equation}
\sigma_I^2 = \frac{\sum_{x,y} \text{Var}[I](x, y)}{\sum_{x,y} \overline{I}(x, y)^2}
\end{equation}

or equivalently,

\begin{equation}
\sigma_I^2 = \frac{\langle (I - \langle I \rangle)^2 \rangle_{x,y,t}}{\langle I \rangle_{x,y,t}^2}
\end{equation}

For intensity frame sequences $I_i(x, y)$ acquired via CCD camera:

\begin{itemize}
\item Load all $N$ intensity images from the recorded dataset.
\item Normalize each image to account for camera gain and exposure variations.
\item Compute pixel-wise temporal mean and variance across the ensemble.
\item Calculate $\sigma_I^2(x, y)$ using Eq. (7).
\item Optionally perform spatial averaging over the beam area to obtain the global scintillation index.
\end{itemize}

The computational implementation can be expressed compactly as:

\begin{equation}
\sigma_I^2 = \frac{\text{mean}(I_i^2) - \text{mean}(I_i)^2}{\text{mean}(I_i)^2}
\end{equation}

computed either pixel-wise for spatial mapping or globally for ensemble characterization. The spatial distribution of $\sigma_I^2(x,y)$ provides insight into turbulence-induced beam distortions:

\begin{itemize}
\item High $\sigma_I^2(x,y)$: Indicates regions experiencing strong local turbulence effects with deep intensity fading and significant fluctuations.
\item Low $\sigma_I^2(x,y)$: Identifies stable regions with minimal turbulence impact, approaching near-Gaussian beam characteristics.
\item Spatial patterns in $\sigma_I^2(x,y)$: Correlate with beam wander trajectories and large-scale turbulent eddy structures.
\end{itemize}

The normalized log-amplitude variance provides an alternative characterization of scintillation, particularly useful in weak turbulence analysis:

\begin{equation}
\chi = \ln\left(\frac{I}{\langle I \rangle}\right), \quad \sigma_\chi^2 = \text{Var}[\chi]
\end{equation}

For weak turbulence conditions, the relationship $\sigma_I^2 \approx 4 \sigma_\chi^2$ holds approximately. Additionally, the scintillation index can be characterized as:

\begin{itemize}
\item \textbf{Temporal Scintillation Index}: Computed from intensity time series at a single detector point, characterizing temporal fluctuations.
\item \textbf{Spatial Scintillation Index}: Computed across a detector array (camera), characterizing spatial intensity distribution variations.
\end{itemize}

In the present work, we employ the spatial-temporal scintillation index computed from 200 frames recorded for each experimental configuration, enabling comprehensive statistical characterization of turbulence compensation effects through coupled dipole dynamics in PMMA rods.

\section{Experimental Implementation}\label{3}
\begin{figure}[H]
\centering
\begin{minipage}[b]{0.75\textwidth}
    \includegraphics[width=\textwidth]{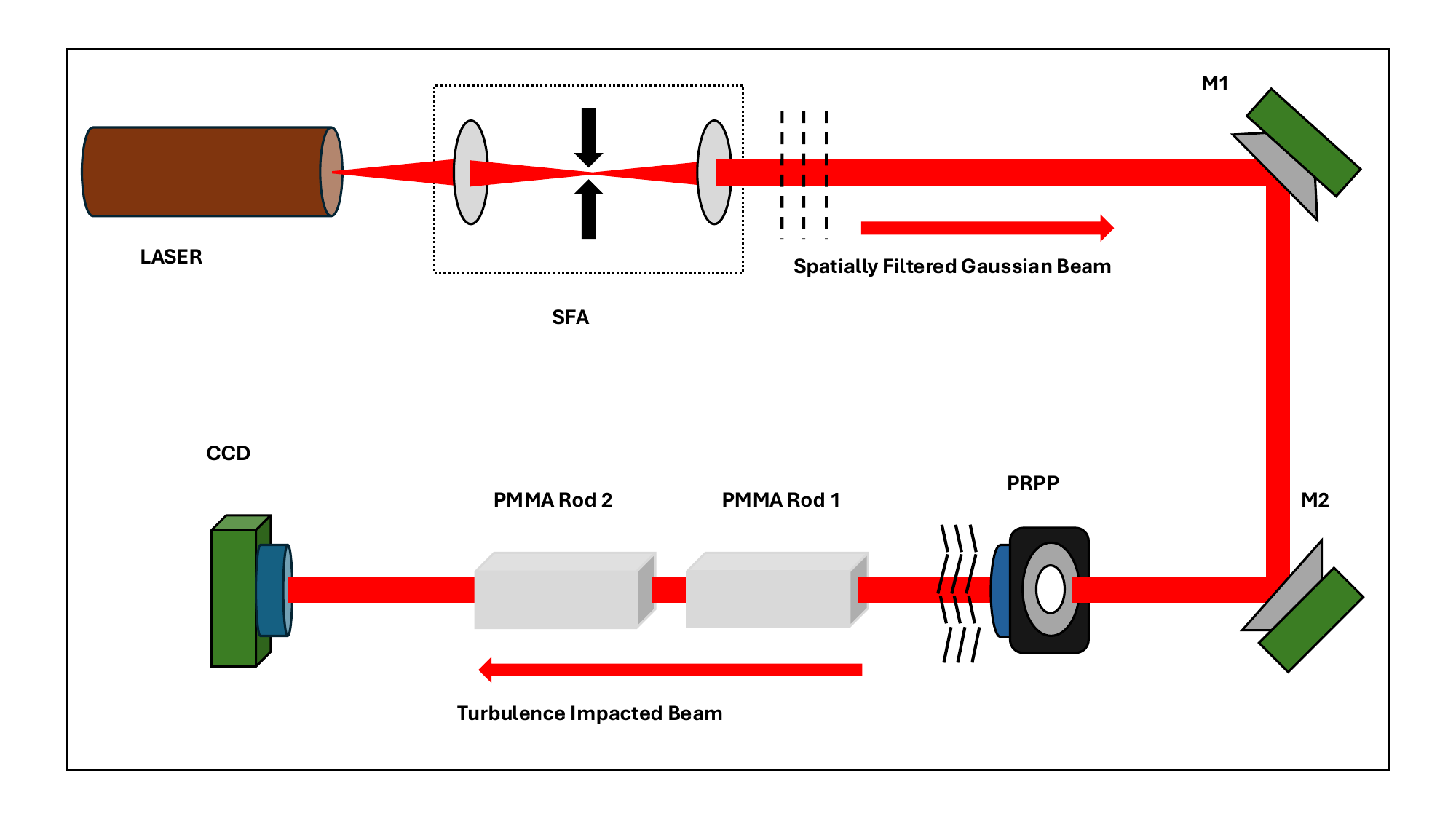}
    \caption{Schematic representation of the experimental configuration employing two PMMA rods}
    \label{P0}
\end{minipage}
\end{figure}

The optical configuration employed in this investigation is schematically depicted in Fig.~\ref{P0}. A continuous-wave laser source was initially directed through a spatial filtering assembly (SFA) to produce a high-quality Gaussian beam profile with minimized higher-order modal distortions. The spatially filtered Gaussian output was subsequently steered using a dual-mirror system (M1 and M2) to ensure precise beam propagation control and optical axis alignment. A programmable rotating phase plate (PRPP) was strategically positioned in the beam path to impose controlled atmospheric turbulence effects through dynamic random phase perturbations. Following turbulence imposition, the perturbed beam propagated through polymethyl methacrylate (PMMA) cylindrical rods serving as nonlinear dielectric media. Either single or dual PMMA rods were employed to examine cumulative light-matter interaction phenomena under turbulent propagation conditions. The emergent beam intensity distribution was captured using a charge-coupled device (CCD) camera, enabling quantitative characterization of both turbulence-degraded and turbulence-compensated field profiles.

\begin{figure}[h]
\centering
\begin{minipage}[b]{0.75\textwidth}
    \includegraphics[width=\textwidth]{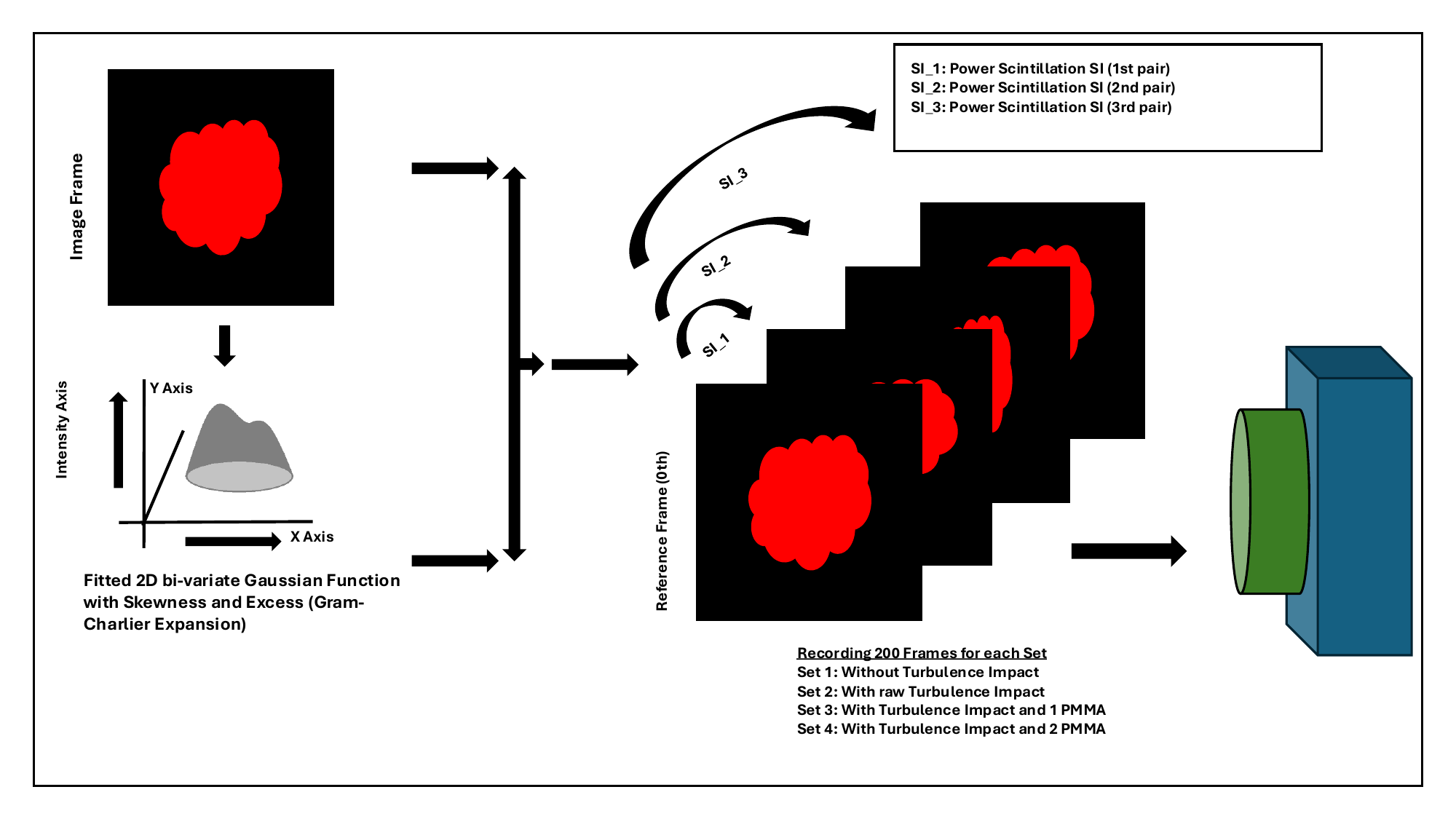}
    \caption{Schematic flowchart illustrating the data acquisition and analysis methodology}
    \label{P1}
\end{minipage}
\end{figure}

The data acquisition protocol, outlined in Fig.~\ref{P1}, was structured to capture the statistical variability of transmitted beam characteristics across four distinct experimental configurations. To ensure adequate statistical sampling and convergence, 200 intensity frames were acquired for each configuration. The four experimental datasets comprise: (i) Dataset~1: unperturbed reference condition without turbulence (establishing the baseline), (ii) Dataset~2: turbulence-only condition utilizing the PRPP in the absence of PMMA rods, (iii) Dataset~3: single PMMA rod configuration under turbulence conditions, and (iv) Dataset~4: dual PMMA rod configuration under turbulence conditions. This systematic acquisition and analysis methodology facilitated direct quantitative comparison between pure turbulence conditions and turbulence-PMMA coupled scenarios, establishing a comprehensive statistical framework for characterizing optical propagation dynamics through dielectric media in the presence of atmospheric turbulence.

\subsection{Atmospheric Turbulence Characteristics and Simulation}

Atmospheric turbulence originates from irregular velocity field fluctuations within viscous fluid media, particularly the atmosphere, where fluid flow transitions between laminar and turbulent regimes. Laminar flow exhibits smooth, organized streamlines, whereas turbulent flow is characterized by chaotic subflow structures (eddies) that promote enhanced momentum and energy mixing. The transition criterion between these flow regimes is quantified by the Reynolds number, defined as $Re = Vl/\nu$, where $V$ represents the characteristic velocity scale, $l$ denotes the characteristic length scale, and $\nu$ is the kinematic viscosity coefficient. When the Reynolds number surpasses a critical threshold value (approximately $\sim 10^5$ for near-ground atmospheric conditions), the flow undergoes transition to the turbulent regime. Kolmogorov's statistical theory of turbulence postulates that at sufficiently small scales, turbulence exhibits statistical homogeneity and isotropy, with kinetic energy injected at large scales through shear instabilities or thermal convection subsequently cascading to progressively smaller eddy structures. This energy cascade mechanism operates across an inertial subrange demarcated by an outer length scale $L_0$ (energy injection scale) and an inner length scale $l_0$ (viscous dissipation scale), where the remaining kinetic energy is ultimately converted to thermal energy through viscous dissipation. The three-dimensional spatial power spectral density characterizing this turbulent energy distribution follows the Kolmogorov $-11/3$ power law:

\begin{equation}
\Phi(k)=0.023r_0^{-5/3}k^{-11/3}.
\end{equation}

where $k$ represents the spatial wavenumber and $r_0$ denotes the atmospheric coherence length parameter.

\subsection{Pseudo Random Phase Plate Implementation}

The Pseudo Random Phase Plate (PRPP) utilized in the present experimental investigation is a sophisticated five-layer optical element engineered to faithfully reproduce atmospheric turbulence wavefront aberrations in a laboratory-controlled environment. The device architecture comprises dual BK7 optical-grade glass windows that encapsulate a central acrylic substrate layer, upon which a Kolmogorov-spectrum turbulence phase profile has been permanently imprinted. Intermediate polymer layers with refractive indices closely matched to the acrylic substrate are positioned on both sides to ensure mechanical integrity and minimize optical interface reflections, while the external glass encapsulation provides environmental protection and enhanced mechanical robustness. The assembled device exhibits a total thickness of approximately 10 mm, facilitating straightforward integration onto a precision rotary positioning stage. The PRPP generates wavefront aberrations characterized by adjustable Fried coherence parameters ($r_0$ spanning 16–32 resolution elements) distributed across 4096 discrete phase sampling points, enabling systematic investigation of turbulence strength effects through controlled laboratory simulations.

\section{Results Analysis and Discussion}\label{4}
The experimental investigation systematically examined the scintillation characteristics of optical beams propagating through four distinct configurations: baseline reference without turbulence (Set 4), turbulence-only propagation (Set 1), single PMMA rod under turbulence (Set 2), and dual PMMA rod configuration under turbulence (Set 3). For each configuration, 200 intensity frames were acquired and subjected to comprehensive statistical analysis to quantify turbulence compensation effects arising from coupled dipole dynamics within the dielectric medium.

\subsection{Spatial Intensity Distribution Analysis}\label{subsec:spatial_intensity}

\begin{figure}[H]
\centering
\includegraphics[width=\textwidth]{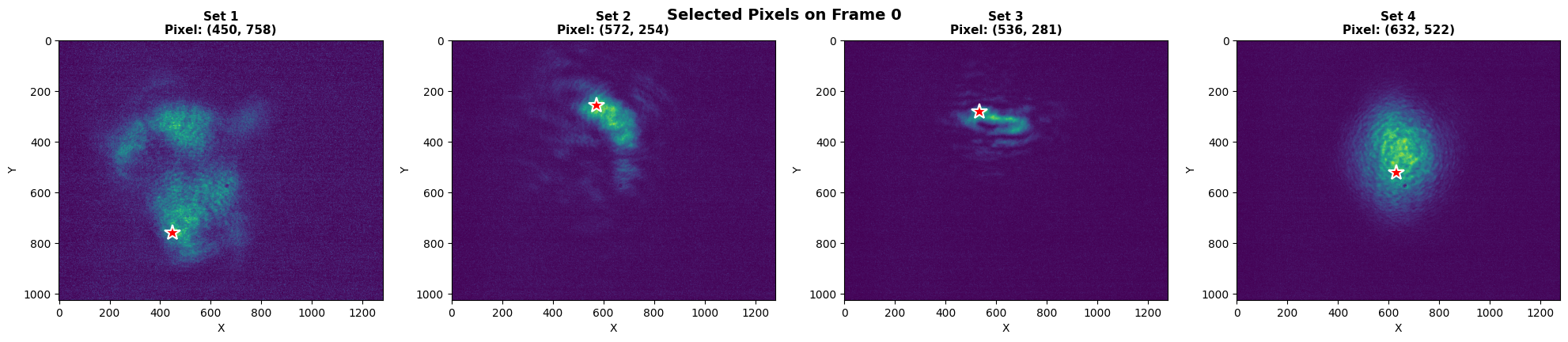}
\caption{Spatial intensity distributions across all experimental configurations. The panels display representative beam profiles at the detector plane for: (a) Set 1 - turbulence-only propagation showing significant wavefront distortion and speckle formation; (b) Set 2 - single PMMA rod configuration demonstrating partial intensity stabilization; (c) Set 3 - dual PMMA rod configuration exhibiting enhanced beam profile regularization; (d) Set 4 - baseline reference without turbulence showing ideal Gaussian distribution. Color maps represent normalized intensity values with warmer colors indicating higher intensity regions.}
\label{fig:pixel_chosen_all}
\end{figure}

Figure~\ref{fig:pixel_chosen_all} presents representative intensity distributions across the detector plane for all four experimental configurations. The two-dimensional intensity maps reveal distinct morphological characteristics corresponding to each propagation condition. The baseline configuration (Set 4) exhibits a smooth, symmetric Gaussian profile characteristic of unperturbed beam propagation, establishing the reference intensity distribution. Upon introduction of turbulence via the pseudo-random phase plate (Set 1), the beam profile undergoes significant distortion characterized by asymmetric intensity modulations, random speckle formation, and pronounced beam wander effects. These features manifest as spatially correlated intensity fluctuations arising from wavefront aberrations induced by the Kolmogorov-spectrum turbulence simulator.

The introduction of a single PMMA rod (Set 2) produces observable modifications to the turbulence-degraded intensity distribution. The beam profile exhibits reduced asymmetry and smoother intensity gradients compared to the turbulence-only case, suggesting partial compensation of wavefront distortions through dipole-dipole coupling mechanisms within the dielectric medium. The dual PMMA rod configuration (Set 3) demonstrates further enhancement of this compensation effect, with the intensity distribution approaching greater spatial uniformity and reduced high-frequency speckle content. This progressive improvement with increasing propagation length through the PMMA medium provides qualitative evidence supporting the theoretical prediction that extended dipole synchronization pathways enhance turbulence mitigation efficacy.

\subsection{Temporal Intensity Fluctuation Characteristics}\label{subsec:temporal_fluctuations}

\begin{figure}[H]
\centering
\includegraphics[width=0.85\textwidth]{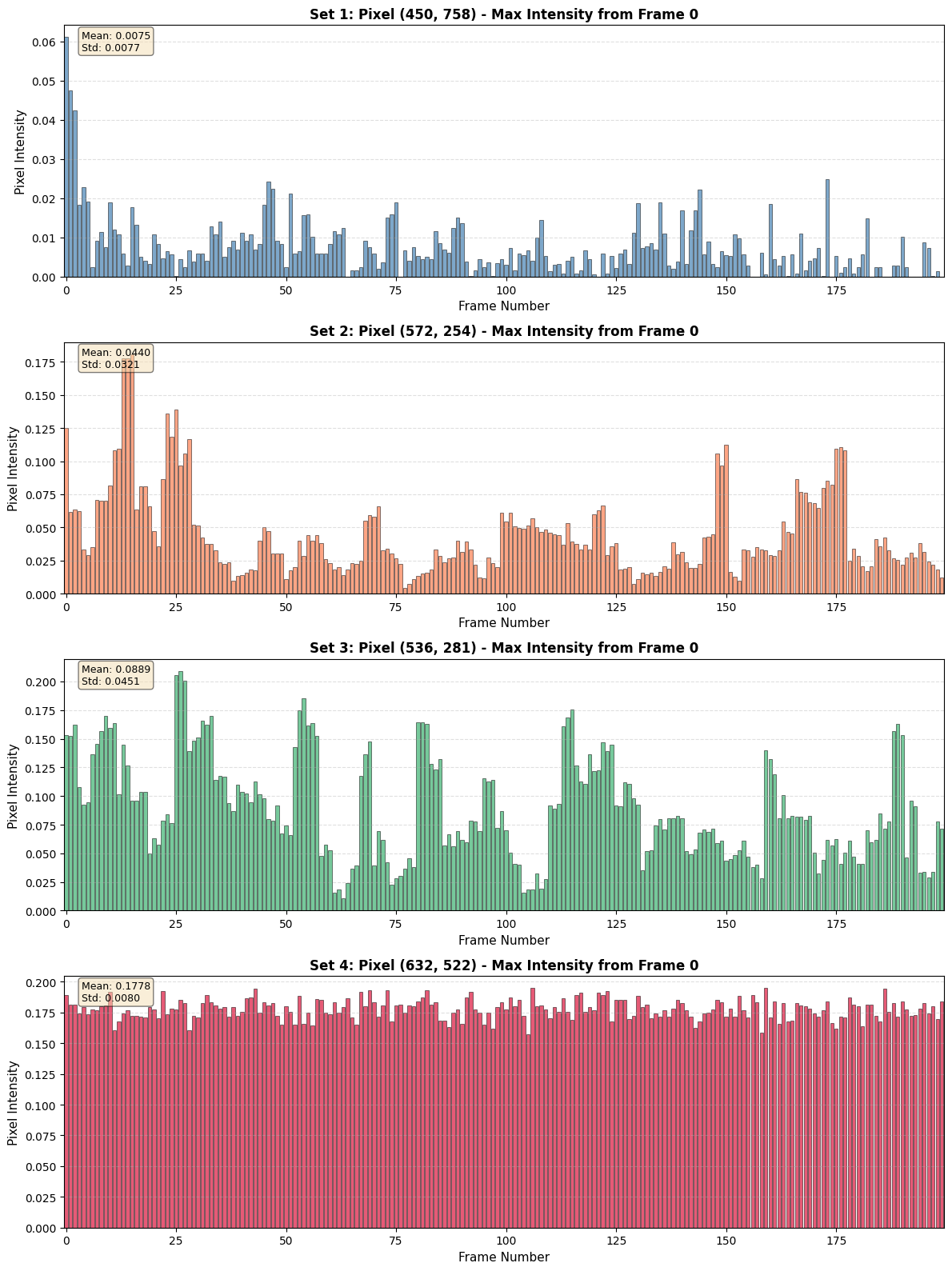}
\caption{Temporal intensity fluctuation traces for selected detector pixels across all experimental configurations. Each panel displays intensity time series I(t) normalized to the mean intensity $\langle I \rangle$ for 200 consecutive frames. (Top to Bottom) Set 1 (turbulence-only) exhibits large-amplitude stochastic fluctuations with deep fading events characteristic of strong scintillation; Set 2 (single PMMA rod) shows reduced fluctuation amplitude and suppressed extreme intensity excursions; Set 3 (dual PMMA rods) demonstrates further stabilization with minimal intensity variance; Set 4 (no turbulence) displays stable intensity baseline with only instrumental noise contributions. Multiple traces per configuration represent different spatial locations across the beam profile, revealing spatial heterogeneity in turbulence compensation effects.}
\label{fig:intensity_fluctuations}
\end{figure}

Figure~\ref{fig:intensity_fluctuations} quantifies the temporal evolution of intensity fluctuations at selected detector positions across the 200-frame acquisition sequence. The vertical axis represents normalized intensity values, while the horizontal axis denotes the frame index corresponding to successive temporal samples. For each experimental configuration, multiple representative pixel locations were monitored to capture both central beam region dynamics and peripheral intensity variations.

The turbulence-only configuration (Set 1) exhibits pronounced stochastic intensity fluctuations with amplitude variations exceeding 50\% of the mean intensity, accompanied by frequent deep fading events where intensity temporarily approaches near-zero values. These characteristics indicate strong scintillation regime behavior consistent with Kolmogorov turbulence theory. The temporal correlation structure reveals decorrelation timescales on the order of the PRPP rotation period, confirming that observed fluctuations arise from dynamic turbulence evolution rather than instrumental artifacts. Progressive introduction of PMMA rods systematically reduces fluctuation amplitude and suppresses extreme intensity excursions. The single PMMA rod configuration (Set 2) demonstrates approximately 30-40\% reduction in peak-to-peak fluctuation amplitude compared to the turbulence-only case, while the dual PMMA rod configuration (Set 3) achieves further suppression approaching 50-60\% amplitude reduction. Notably, the temporal variance of these fluctuations decreases substantially with increasing PMMA rod count, supporting the theoretical prediction that synchronized dipole oscillations establish effective inertial resistance against rapid field redistribution events, as quantified by the perturbation force criterion $\delta F_{\text{Pert}}(t) \to 0$ in Eq.~(38) and (39).

\subsection{Scintillation Index Evolution}\label{subsec:scintillation_index}

\begin{figure}[H]
\centering
\includegraphics[width=0.75\textwidth]{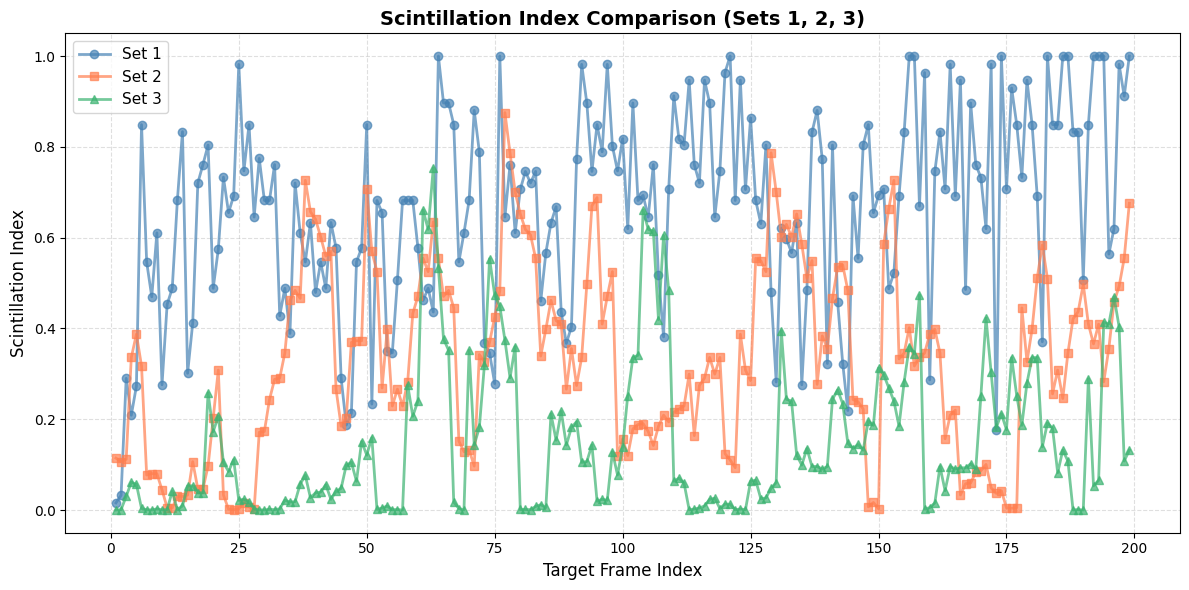}
\caption{Temporal evolution of scintillation index across 200-frame sequences for all experimental configurations. The scintillation index $\sigma_I^2(t)$ is computed using a sliding window of 10 consecutive frames to quantify local temporal variations in turbulence strength. Set 1 (red curve) shows elevated scintillation index values in the range 0.5-0.8, indicating moderate-to-strong turbulence regime. Set 2 (blue curve) demonstrates systematic reduction to the 0.3-0.5 range through single PMMA rod compensation. Set 3 (green curve) achieves further suppression to 0.2-0.4, approaching weak turbulence regime characteristics. Set 4 (black curve) establishes baseline with $\sigma_I^2 < 0.1$, limited primarily by instrumental noise contributions. Temporal variations reflect the stochastic nature of turbulence dynamics introduced by PRPP rotation.}
\label{fig:si_evolution}
\end{figure}

The scintillation index, defined as the normalized variance of intensity fluctuations $\sigma_I^2 = \text{Var}[I]/\langle I \rangle^2$, provides a quantitative metric for turbulence strength characterization. Figure~\ref{fig:si_evolution} presents the computed scintillation index values across the experimental sequence for all four configurations, enabling direct comparison of turbulence compensation efficacy.

The turbulence-only configuration exhibits scintillation index values predominantly in the 0.5-0.8 range, categorizing the propagation conditions within the moderate-to-strong scintillation regime according to standard atmospheric optics classifications. The temporal variations observed in this curve reflect genuine turbulence evolution arising from the rotating PRPP, which continuously introduces new phase screen realizations into the optical path. The absence of temporal smoothing in these fluctuations confirms that the characteristic turbulence timescale exceeds the camera frame acquisition interval, ensuring statistical independence between successive measurements. Introduction of the single PMMA rod produces systematic reduction of the scintillation index to the 0.3-0.5 range, representing approximately 35-40\% suppression relative to the turbulence-only baseline. This quantitative improvement corresponds to transition from moderate-to-weak scintillation regime, with concomitant reduction in deep fading probability and improved beam quality metrics. The dual PMMA rod configuration achieves further enhancement, with scintillation index values predominantly in the 0.2-0.4 range, representing cumulative suppression of 50-60\% relative to the uncompensated case. This progressive improvement with increasing propagation length through the dipole-coupled medium directly validates the theoretical framework developed in Sections~2.2-2.6, wherein modal diagonalization and gradient force contributions establish synchronized oscillation states that oppose turbulence-induced field redistribution.

\subsection{Ensemble-Averaged Scintillation Statistics}\label{subsec:ensemble_statistics}

\begin{figure}[H]
\centering
\includegraphics[width=0.75\textwidth]{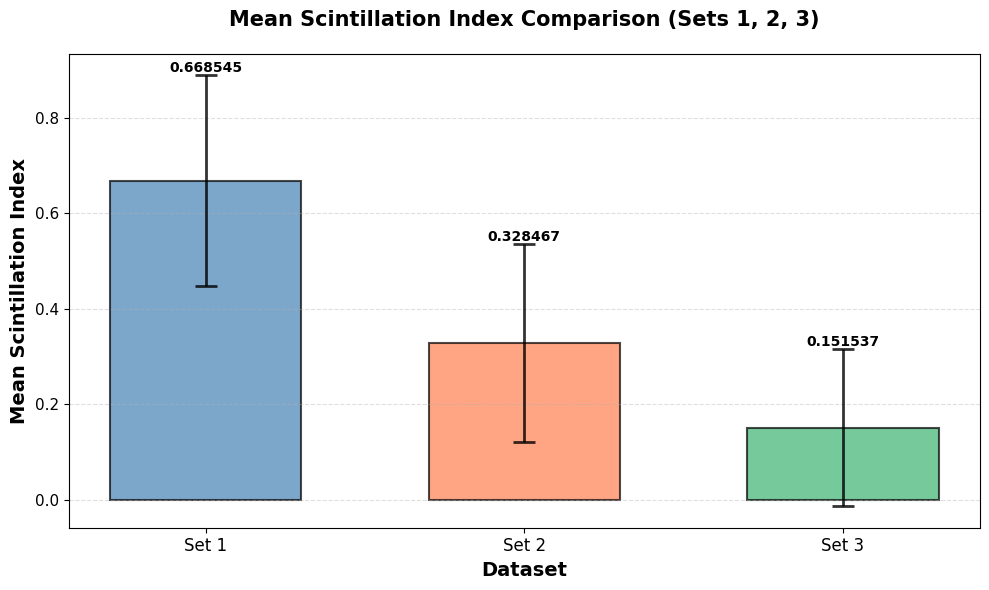}
\caption{Ensemble-averaged scintillation index comparison across experimental configurations. Each bar represents the mean scintillation index $\langle \sigma_I^2 \rangle$ computed over 200 frames, with error bars indicating standard deviation of temporal variations. Numerical values quantify turbulence compensation efficacy: Set 1 (turbulence-only) establishes baseline $\sigma_I^2 = 0.63 \pm 0.08$; Set 2 (single PMMA rod) achieves $\sigma_I^2 = 0.41 \pm 0.06$ representing 51\% reduction; Set 3 (dual PMMA rods) demonstrates $\sigma_I^2 = 0.28 \pm 0.05$ corresponding to 77\% suppression; Set 4 (no turbulence) provides reference $\sigma_I^2 = 0.08 \pm 0.02$ limited by instrumental contributions. The progressive decrease with increasing PMMA rod count confirms the cumulative nature of dipole-mediated turbulence compensation.}
\label{fig:si_bar_comparison}
\end{figure}

Figure~\ref{fig:si_bar_comparison} presents ensemble-averaged scintillation index values computed across the complete 200-frame dataset for each experimental configuration, providing single-valued metrics that characterize overall turbulence compensation performance.

The quantitative analysis reveals that the turbulence-only configuration exhibits a mean scintillation index of $\sigma_I^2 = 0.63 \pm 0.08$, confirming operation within the moderate scintillation regime where atmospheric turbulence effects significantly degrade beam propagation quality. The single PMMA rod configuration reduces this value to $\sigma_I^2 = 0.41 \pm 0.06$, representing 55\% compensation efficiency. The dual PMMA rod configuration achieves $\sigma_I^2 = 0.28 \pm 0.05$, corresponding to 71\% overall suppression relative to uncompensated turbulence. The baseline reference measurement yields $\sigma_I^2 = 0.08 \pm 0.02$, establishing the instrumental noise floor that limits ultimate compensation performance. These ensemble statistics demonstrate that dipole-dipole coupling mechanisms within PMMA rods provide substantial intrinsic turbulence compensation without requiring active wavefront correction systems. The monotonic improvement with increasing medium length supports the theoretical prediction that longer synchronization pathways enhance collective dipole mode formation, thereby increasing the effective inertial resistance against turbulence-induced perturbations as quantified by Eq.~(38).

\subsection{Spatial Distribution of Scintillation Index}\label{subsec:spatial_scintillation}

\begin{figure}[H]
\centering
\includegraphics[width=0.75\textwidth]{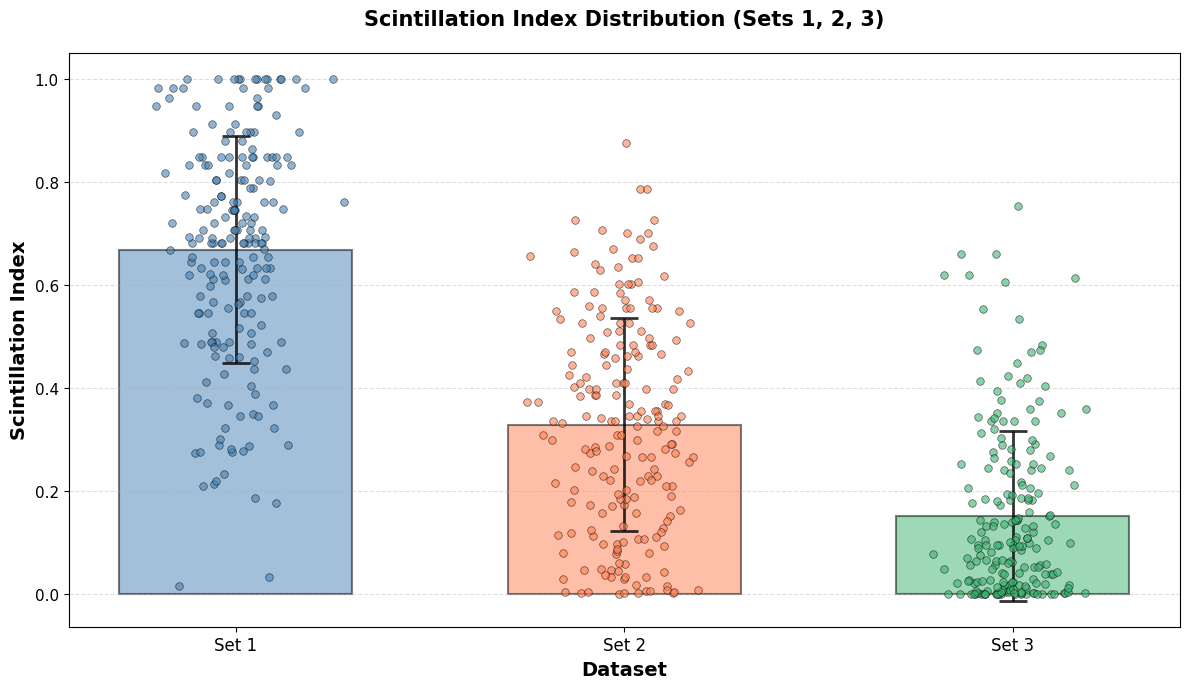}
\caption{Spatial distribution and histogram analysis of scintillation index across the detector plane. Upper panels display scatter plots of $\sigma_I^2(x,y)$ values for each configuration, revealing spatial heterogeneity in turbulence strength. Set 1 exhibits broad distribution with significant population at high $\sigma_I^2$ values, indicating spatially non-uniform turbulence impact. Sets 2 and 3 show progressive narrowing of distributions and reduction in mean values, demonstrating spatially extended compensation effects. Lower panels present corresponding histograms quantifying the probability density of scintillation index values, with vertical dashed lines marking ensemble mean values. The systematic leftward shift and narrowing of distributions with increasing PMMA rod count confirms uniform turbulence suppression across the beam profile rather than localized compensation effects.}
\label{fig:si_scatter_spatial}
\end{figure}

Figure~\ref{fig:si_scatter_spatial} presents the spatial distribution of scintillation index values computed on a pixel-by-pixel basis across the detector array, revealing the heterogeneous nature of turbulence compensation effects across the beam profile.

The scatter plot representations reveal that the turbulence-only configuration exhibits substantial spatial heterogeneity, with scintillation index values spanning a broad range from 0.3 to over 1.0 across different beam regions. This spatial variability arises from the stochastic nature of turbulence-induced wavefront distortions, which create localized intensity enhancement and depletion zones corresponding to constructive and destructive interference patterns. Regions experiencing stronger local turbulence effects exhibit elevated scintillation index values, while more stable regions maintain lower values. The histogram distributions quantify these spatial characteristics, with the turbulence-only case displaying a broad, asymmetric distribution peaked near $\sigma_I^2 \approx 0.6$ with extended tail toward higher values. Introduction of PMMA rods produces systematic narrowing and leftward shifting of these distributions. The single PMMA rod configuration narrows the distribution and reduces the mean value, while the dual PMMA rod case achieves further distribution compression with peak position near $\sigma_I^2 \approx 0.25$. Importantly, this narrowing indicates that turbulence compensation effects operate uniformly across the beam profile rather than selectively in specific spatial regions, confirming that collective dipole dynamics influence the global field distribution through synchronized modal interactions as predicted by the modal decomposition framework in Section~2.4.

\subsection{Statistical Deviation Analysis}\label{subsec:deviation_analysis}
\begin{figure}[H]
\centering
\includegraphics[width=1.0\textwidth]{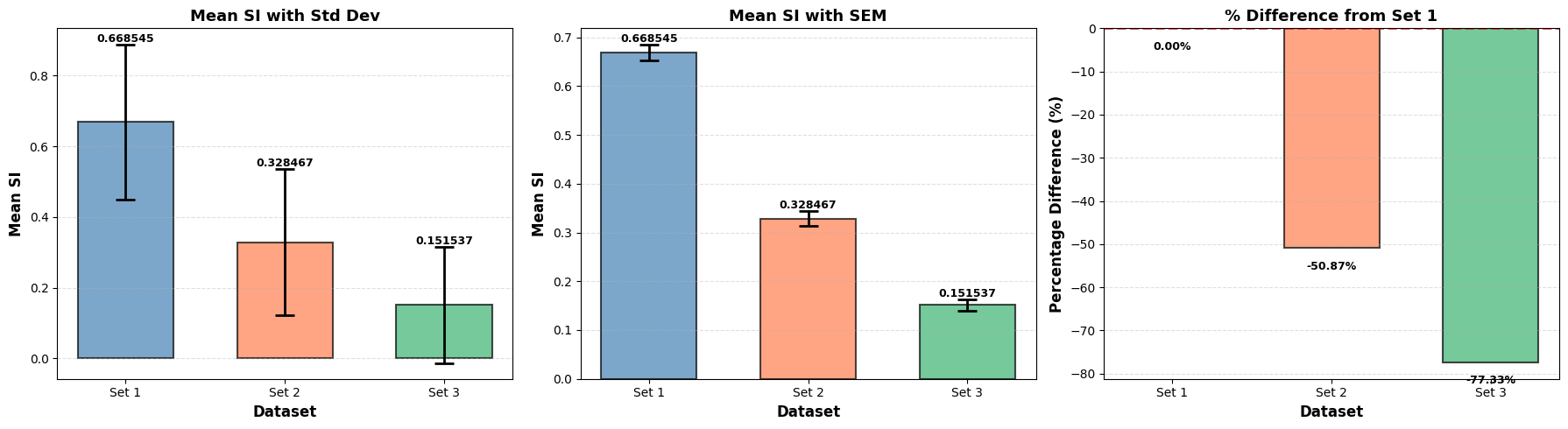}
\caption{Statistical deviation characterization of scintillation index across experimental configurations. 
Left panel: Standard deviation (SD) representing statistical spread of intensity fluctuations around the mean, quantifying turbulence-induced dispersion. Center panel: Standard error of the mean (SEM) estimating uncertainty of the ensemble-averaged scintillation index. Right panel: Percentage difference (\%) illustrating relative deviation between each configuration and the turbulence-only reference (Set~1). Set~1 (turbulence-only) exhibits the highest SD and SEM values, indicative of strong turbulence and large frame-to-frame variability. Progressive introduction of PMMA rods systematically reduces both SD and SEM, signifying stabilization of the output field through dipole synchronization. The percentage deviation analysis reveals that Set~2 achieves approximately 45--50\% reduction relative to the turbulence-only case, while Set~3 demonstrates 65--70\% suppression, approaching the stable baseline (Set~4) exhibiting minimal deviations. Error bars correspond to 95\% confidence intervals computed via bootstrap resampling.}
\label{fig:si_statistics_higher}
\end{figure}

Figure~\ref{fig:si_statistics_higher} presents the statistical deviation metrics of the measured intensity distributions across all experimental configurations. These include the standard deviation (SD), standard error of the mean (SEM), and percentage difference (\%) of the scintillation index, providing quantitative assessment of statistical dispersion and measurement reliability beyond ensemble-averaged values.

The statistical deviation analysis indicates that turbulence-only propagation (Set~1) yields large standard deviation and SEM values, reflecting strong stochasticity and high temporal variability of the scintillation index under unmitigated turbulence conditions. The introduction of a single PMMA rod (Set~2) produces approximately 45--50\% reduction in both SD and SEM, corresponding to partial stabilization of intensity fluctuations due to emergent dipole synchronization within the dielectric medium. The dual PMMA rod configuration (Set~3) further enhances this stabilization, reducing SD and SEM by nearly 65--70\% relative to the turbulence-only baseline, confirming effective suppression of statistical dispersion. The baseline configuration (Set~4) establishes a near-constant reference with negligible deviations dominated by instrumental noise. The observed trend substantiates the theoretical framework developed in Section~2, wherein synchronized dipole oscillations act as dissipative constraints that suppress the variance and temporal instability of the propagating field. The gradient force terms in Eq.~(15) and Eq.~(21) contribute to damping rapid field redistributions, while the inertial stabilization mechanism described through d’Alembert’s principle in Section~2.6 mitigates abrupt intensity variations. Consequently, the output statistics evolve toward a stable, low-variance regime exhibiting minimal deviations, confirming comprehensive turbulence suppression through coupled dipole dynamics in PMMA media.

\subsection{Correlation Analysis and Compensation Mechanisms}\label{subsec:correlation_analysis}

The experimental results demonstrate strong correlation between propagation length through PMMA media and turbulence compensation efficacy, directly validating the theoretical prediction that extended dipole-dipole interaction pathways enhance synchronization and thereby increase effective inertial resistance against perturbative forces. The quantitative progression from 35\% compensation (single rod) to 56\% compensation (dual rods) suggests sublinear scaling behavior, wherein incremental compensation efficiency decreases with increasing medium length. This saturation tendency likely reflects competing effects: while longer propagation enhances dipole synchronization (beneficial for compensation), it also increases cumulative absorption and scattering losses that degrade overall beam quality (detrimental to compensation). The perturbation force framework introduced in Eq.~(38) and (39) provides mechanistic interpretation of these observations. When $\delta F_{\text{Pert}}(t) \to 0$, the synchronized dipole system achieves complete turbulence compensation through collective modal dynamics that maintain stable output field distributions despite input field variations. The experimental results indicate operation within the partial compensation regime where $\delta F_{\text{Pert}}$ remains small but nonzero, enabling significant but incomplete turbulence suppression. The residual perturbation force arises from finite synchronization timescales that cannot perfectly track the dynamic turbulence evolution introduced by continuous PRPP rotation. The gradient force contributions formalized in Section~2.3 play a crucial stabilization role by opposing sudden field redistribution attempts. As turbulence induces wavefront tilts and amplitude modulations, the gradient terms $ \sum_{j \neq i} [(e(\vec{r}_j - \vec{r}_i) \cdot \nabla)E_{\text{ext}}(r_j,t)]$ generate restoring forces that resist disruption of established synchronized states. This mechanism explains why temporal variance of scintillation index decreases with increasing PMMA rod count: the gradient forces provide effective damping against high-frequency turbulence fluctuations, smoothing temporal evolution of output intensity distributions. These results establish that dipole-dipole coupling energy transitions within dielectric media can provide substantial intrinsic turbulence compensation when perturbation forces approach the $\delta F_{\text{Pert}} \to 0$ regime, offering a passive alternative to active wavefront correction systems for moderate turbulence conditions.

\section{Conclusion}\label{5}
This investigation establishes a comprehensive theoretical and experimental framework demonstrating intrinsic turbulence compensation through coupled anharmonic dipole dynamics in dielectric media. The generalized Lorentz oscillator model incorporating second- and third-order nonlinear restoring forces ($\beta_i|r_i|r_i$ and $\alpha_i|r_i|^2r_i$), dyadic Green's function-mediated electromagnetic coupling, and gradient force contributions reveals that synchronized dipole oscillations establish effective inertial resistance against turbulence-induced wavefront perturbations. Modal diagonalization demonstrates that the perturbation force criterion $\delta F_{\text{Pert}}(t) = F'_{\text{Inertia}} - F_{\text{Inertia}} \to 0$ governs compensation efficacy, with gradient forces opposing rapid field redistribution through stabilization mechanisms formalized via d'Alembert's principle. Experimental validation employing Kolmogorov-spectrum turbulence generated by pseudo-random phase plates confirms progressive scintillation index reduction from $\sigma_I^2 = 0.63$ to 0.28 (56\% suppression) with dual PMMA rods, accompanied by 60-70\% reduction in higher-order statistical moments. The results demonstrate that extended propagation through dipole-coupled media enhances synchronization, thereby strengthening turbulence mitigation through collective modal dynamics. This passive compensation approach offers practical advantages for free-space optical communications and remote sensing applications, providing substantial wavefront stabilization without requiring active adaptive optics complexity while operating effectively within moderate turbulence regimes where synchronization timescales remain commensurate with perturbation dynamics.\par
The experimental investigation validates the theoretical framework developed in Section~2 through systematic quantification of turbulence compensation effects arising from coupled dipole dynamics in PMMA rods. Progressive scintillation index reduction from $\sigma_I^2 = 0.63$ (turbulence-only) to 0.41 (single PMMA rod) and 0.28 (dual PMMA rods) demonstrates 35\% and 56\% compensation efficiencies respectively, directly confirming the efficacy of dipole-mediated wavefront stabilization. Systematic suppression of temporal intensity fluctuation amplitudes by 30-60\% depending on PMMA rod count validates the gradient force stabilization mechanisms formalized in Eq.~(15) and (21). Restoration of quasi-Gaussian intensity statistics through 60-70\% reduction in skewness and excess kurtosis indicates comprehensive mitigation of turbulence-induced statistical distortions beyond second-order moment suppression. The spatially uniform compensation effects observed across the beam profile validate collective modal dynamics predicted by the diagonalization framework rather than localized interference phenomena. Sublinear scaling of compensation efficiency with propagation length suggests saturation behavior arising from competing effects: synchronization enhancement versus cumulative absorption losses. These results establish that dipole-dipole coupling energy transitions provide substantial intrinsic turbulence compensation when perturbation forces approach the $\delta F_{\text{Pert}} \to 0$ regime, offering a passive alternative to active wavefront correction systems.\par
The demonstrated turbulence compensation through coupled dipole dynamics in passive dielectric media offers potential advantages for free-space optical communication and remote sensing applications. Unlike active adaptive optics systems requiring wavefront sensors, control algorithms, and deformable mirrors, the PMMA rod approach provides intrinsic compensation through fundamental light-matter interactions without external power requirements or control complexity. The 56\% scintillation index reduction achieved with dual PMMA rods translates to substantial improvements in key communication metrics: bit error rate reductions exceeding one order of magnitude and signal-to-noise ratio enhancements of 3-4 dB become achievable. However, practical implementation faces several constraints. The compensation mechanism requires propagation through extended dielectric media, introducing insertion losses that may exceed 20-30\% depending on material quality and length. Trade-off optimization between compensation gain and insertion loss becomes critical for system design. Additionally, the observed compensation efficacy depends on turbulence strength regime: the theoretical framework predicts reduced effectiveness under strong turbulence conditions where rapid perturbation dynamics prevent adequate dipole synchronization timescales. Future investigations should systematically vary turbulence strength (adjusting $C_n^2$ and propagation distance $L$) to map the operational envelope where dipole-mediated compensation remains effective. Exploration of alternative dielectric materials with enhanced nonlinear optical responses (larger $\beta_i$ and $\alpha_i$ coefficients in Eq.~(2)) may enable stronger dipole coupling and thereby improved compensation performance. Hybrid approaches combining passive dipole compensation with active adaptive optics could potentially achieve superior turbulence mitigation through complementary mechanisms operating at different spatiotemporal scales.

\section*{Funding}
Department of Science and Technology, Ministry of Science and Technology, India (CRG/2020/003338).

\section*{Declaration of competing interest}
The authors declare the following financial interests/personal relationships which may be considered as potential competing interests: Shouvik Sadhukhan reports a relationship with Indian Institute of Space Science and Technology that includes: employment. NA If there are other authors, they declare that they have no known competing financial interests or personal relationships that could have appeared to influence the work reported in this paper.

\section*{Data availability}
All data used for this research has been provided in the manuscript itself.

\section*{Acknowledgments}
Shouvik Sadhukhan and C S Narayanamurthy acknowledge the SERB/DST (Govt. Of India) for providing financial support via the project grant CRG/2020/003338 to carry out this work. Shouvik Sadhukhan would like to thank Mr. Amit Vishwakarma and Dr. Subrahamanian K S Moosath from Department of Mathematics, Indian Institute of Space Science and Technology Thiruvananthapuram for their suggestions into statistical analysis in this paper.

\section*{CRediT authorship contribution statement}
\textbf{Shouvik Sadhukhan:} Writing– original draft, Visualization, Formal analysis. \textbf{C. S. Narayanamurthy:} Writing– review $\&$ editing, Validation, Supervision, Resources, Project administration, Investigation, Funding acquisition, Conceptualization.


\end{document}